# A Remote Carrier Synchronization Technique for Coherent Distributed Remote Sensing Systems

Juan Carlos Merlano-Duncan, *Senior Member, IEEE*, Liz Martinez-Marrero, *Student, IEEE*, Jorge Querol, *Member IEEE*, Sumit Kumar, *Member, IEEE*, Adriano Camps, *Fellow, IEEE*, Symeon Chatzinotas, *Senior Member, IEEE*, Björn Ottersten, *Fellow, IEEE*

*Abstract*—Phase, frequency, and time synchronization are crucial requirements for many applications, such as multi-static remote sensing and communication systems. Moreover, the synchronization solution becomes even more challenging when the nodes are orbiting or flying on airborne or spaceborne platforms. This paper compares the available technologies used for the synchronization and coordination of nodes in distributed remote sensing applications.

Additionally, this paper proposes a general system model and identifies preliminary guidelines and critical elements for implementing the synchronization mechanisms exploiting the inter-satellite communication link. The distributed phase synchronization loop introduced in this work deals with the self-interference in a full-duplex point to point scenario by transmitting two carriers at each node. All carriers appear with different frequency offsets around a central frequency, called the application central-frequency or the beamforming frequency. This work includes a detailed analysis of the proposed algorithm and the required simulations to verify its performance for different phase noise, AWGN, and Doppler shift scenarios.

*Index Terms*—Synchronization, multi-static remote sensing systems, distributed beamforming, Phase-Locked-Loops.

## I. INTRODUCTION

Distributed payloads, decentralized systems, cooperative platforms, and collaborative beamforming are crucial elements enabling the next generation of multi-static remote sensing systems. Some examples of distributed remote sensing applications are the bistatic and multi-static SAR [1]. A recent example is the European Space Agency Harmony Mission (within the Earth Explorer 10 program), in which two identical receive-only spacecraft follow Sentinel-1D in a formation, and use it as a radar illuminator [2]. On the other hand, multistatic configurations are the only feasible alternatives to achieve the radar power budget in missions from MEO/GEO orbits [3]–[6].

Similarly, for microwave radiometry applications, the distributed and formation flying configurations would be a game-changing technology [7]. The spatial resolution of a single platform radiometer can be improved only by increasing its aperture size. Therefore, the use of formation flying configurations provides the potential to increase spatial resolution significantly [8]–[13]. This technique can also be applied in 3D synthetic aperture radiometers [14], [15], in contrast to 2D coplanar arrays, providing the system with more flexibility and giving the possibility to diminish the mutual coupling between the antennas.

All these new configurations have stringent requirements in terms of absolute phase, frequency, and time synchronization. Since the signal generation is performed locally at each distributed node, achieving precise synchronization is a very challenging task. The synchronization becomes even more difficult when the geometric distance between the distributed nodes is considerable in terms of the signal wavelengths and, in particular, when this electrical distance is time-varying due to changes in the conditions of the transmission medium. This effect is observed for the case of nodes flying, hovering, or orbiting on aerial or space platforms.

The literature on distributed beamforming has been quickly populated during the last decade, and different theoretical and analytical models, algorithms, and techniques have emerged during the last years. Nevertheless, there are not many practical implementations, even at a research stage. The main limiting factor to make practical implementation feasible is the synchronization under realistic scenarios. In these cases, the requirements in terms of implementation effort, power consumption, and complexity needed to achieve the synchronization goals may surpass the ones required by the primary payload application itself.

Some practical solutions have been proposed in the mobile communications area, but the same technical challenges mentioned above were also an impediment for actual implementations. These impediments triggered the development of other Multiple-Input-Multiple-Output (MIMO) solutions with centralized synchronization relaying into

Manuscript submitted March 31$^{st}$, 2020. This work was supported by the Fond National de la Recherche Luxembourg, under the CORE projects COHESAT: Cognitive Cohesive Networks of Distributed Units for Active and Passive Space Applications, and 5G-SKY. This paper was presented in partially at IGARSS 2019.

J.C. Merlano-Duncan, L. Martinez-Marrero, J. Querol, S. Kumar, S. Chatzinotas and B. Ottersten are with the Interdisciplinary Centre for Security Reliability and Trust (SnT), University of Luxembourg, 1855 Luxembourg City, Luxembourg (e-mail: juan.duncan@uni.lu; liz.martinez-marrero@uni.lu; jorge.querol@uni.lu, sumit.kumar@uni.lu, symeon.chatzinotas@uni.lu; bjorn.ottersten@uni.lu).

Adriano Camps is with the Remote Sensing Lab, Universitat Politecnica de Catalunya and UPC/IEEC, 08034 Barcelona, Spain (e-mail: camps@tsc.upc.edu)

backhauling-network links or GNSS signals instead of wireless synchronization techniques embedded in the communications standard. Besides, the synchronization backhauling via wires is not possible in all practical cases.

The field of Wireless Sensor Networks (WSN) has also attracted interest in distributed beamforming and remote clock synchronization due to rapid development in sensor technologies and embedded systems using low power equipment. Since WSN deals with weak power signals, the distributed and collaborative beamforming looks more appealing than in other types of wireless networks, in principle by the natural topology of the sensor networks, and also by its power constraints [16].

Several other application fields require synchronization of distributed radio systems, such as, for instance, very large baseline phase arrays. In the last years, several synchronization techniques for telecommunications using distributed radio systems have been proposed. In 1968, Thompson et al. [17] discussed, compared, and classified the available methods for reducing propagation-induced phase fluctuations in frequency distribution systems and defined the principle of round trip stabilization systems. These works were usually applied to very long baseline arrays for radio astronomy applications. This last case may also be seen as a distributed radio system, but with the nodes in a fixed static position. In these cases, the signal distribution and the remote synchronization is performed using an auxiliary wired or wireless transmission channel. The phase synchronization techniques described in [17] can be seen as a sort of distributed Phase Locked Loop (PLL). In those cases, a direct feedback action is applied over a two-way transmission media as part of a whole phase loop under the assumption of channel reciprocity. Even in this case, where the radio units have a fixed position and are interconnected with coaxial cables, the design is very challenging due to the persistent random variations of the electrical length of the transmission media used for the synchronization. These phase fluctuations come as a function of the temperature in the cables and other physical parameters that cannot be easily characterized. The synchronization becomes even more challenging when one or more nodes are constantly moving and a wireless media is used for the synchronization link. The main problems found in this case, on top of the mechanical movement of the nodes, are fading, multi-path and non-reciprocity of the channel, which makes a practical implementation of multi-static systems almost impossible in cluttered or indoor radio environments. Another challenge in distributed remote sensing systems is the trade-off between the radio resources, such as power and spectrum, used by the synchronization and coordination mechanism in comparison to the resources used by the sensing process itself.

Under these constraints, a set of scenarios for which distributed microwave remote sensing can flourish in future practical applications can be foreseen. The main factor that will benefit the implementation of distributed coherent sensing will be the availability of a cost-efficient inter-node communication channel suitable for synchronization and coordination. A good example of these scenarios is the one that comprises swarms or formation flying topologies implemented using spaceborne or airborne platforms.

In this paper, we study and compare the available technologies used for synchronization and coordination of nodes in distributed remote sensing applications. Additionally, we propose a dual-carrier remote phase synchronization system and identify preliminary guidelines and critical elements for the implementation of the synchronization mechanism.

This manuscript is organized as follows: Section II presents the system model to be used during the assessment. Section III contains a classification of the different synchronization systems. Section IV describes some of the examples of remote clock synchronization already present in the literature. Section V presents the drawbacks of the single-frequency in-band full-duplex synchronization loop. Section VI shows the analysis of the proposed dual carrier synchronization loop. Section VII shows the simulations of the dual carrier synchronization loop. Finally, Section VIII contains the conclusions of this manuscript.

II. SYSTEM MODEL

A typical remote sensing system consists of distributed wireless radio nodes. Nodes operate their remote sensing operations independently; however, they cooperate to perform beamforming operations. In general, the distributed remote sensing deployment can have different classifications according to the topology, nodes mobility, or how they perform the synchronization. The last one of them is the most interesting for this work. Specifically, for synchronization purposes, distributed remote sensing systems can be classified as follows:

- Systems where the synchronization relies on the wired distribution of a common reference signal, such as the Very Large Array (VLA) used in Radio astronomy.
- Systems that depend on wireless communications to synchronize the distributed sensors. For example, High Altitude Platform Systems (HAPS), stratospheric weather balloons, and satellite constellations. Our work is focused on this group.

The target applications for remote sensing, depending on the sensor type, can be either passive or active. Passive sensor nodes can perform only receive beamforming. However, with the active nodes, both transmit and receive beamforming can be performed. Nonetheless, both the active and passive nodes are equipped with transmit and receive modules in order to achieve synchronization, which is required to perform multi-beam beamforming.

In our system model, a distributed deployment of N radio nodes, i.e., no particular topology is considered. Each of them is associated with a state variable. Additionally, each node can communicate with other nodes in the array. All the variables are a function of time and frequency; hence, for the sake of simplicity, we do not use the time and frequency variables in the following representations. Let the channel matrix between the nodes denoted by $H$, where $h_{nm} \in H$ means the single tap channel for the waveform transmitted from the $n^{th}$ radio towards the $m^{th}$ radio. The channel response varies with time, altitude, and the radio frequency hardware. A typical schematic of the distributed sensing system is illustrated in Fig. 1.

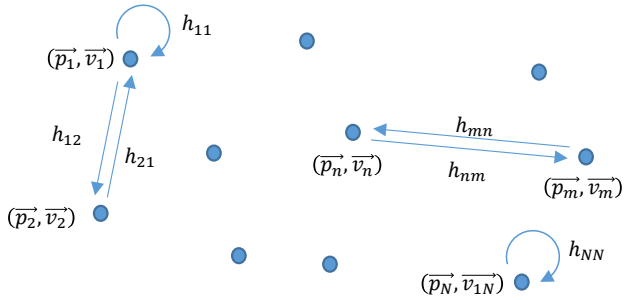

Fig. 1. Distributed set of radio nodes intended to perform a remote sensing system. The radio nodes are responsible for synchronization and beamforming. The elements $h_{11}$ to $h_{NN}$ are the self-interference elements that appear in an in-band full duplex channel interconnection. The self-interference signal reception can be used for self-calibration and self-tracking of parameters such as phase and timing offsets. $\vec{p_n}$ and $\vec{v_n}$ denote position and velocity of the nth node.

Since all the radio nodes are capable of autonomous operation, they generate their own initial time and phase reference signal. However, none of the references is equal to each other. The phases and drifts are also different to the extent that no coherent processing can be performed.

A synchronization among them is compulsory, the purpose of which is to establish a lock on initial time and phase reference signals with respect to a master reference. The master reference is selected to be the one with the best frequency stability and phase noise performance. To establish such a reference lock, every combination of the radio nodes (**n,m**) transmit a waveform modulated with a known and controlled amplitude and phase[18].

Allowing every node to have a possible communication with other nodes will increase the SNR and, therefore, the quality of the synchronization of each node while avoiding the challenging requirements of big directive antennas and their accurate pointing. This mechanism will allow using all the elements in the array to disseminate the frequency reference given by the master node. These capabilities come at the cost of power and spectral resources. These resources can be split among the nodes in time, frequency, and code. For instance, if the DVB-S2X standard is used for this purpose, each node phase is only transmitted in a pilot field, and additionally, the waveforms from one node to the other nodes are multiplexed using different Walsh-Hadamard codes. Using this approach, an array of tens of nodes will require splitting the spectrum into tens of sub-carriers. Furtherly, for very dense arrays with hundreds or even more nodes, it will be reasonable that after a limited number of nodes (tens) operate in closed-loop synchronization, most of them synchronize passively to the array using the available signals.

### III. SYSTEM CLASSIFICATION

A synchronization system for a distributed remote sensing application can be classified on the basis of radio resource allocation (frequency bands) for performing inter-node communication. There are three main classes, as described below and represented in Fig. 2:
- Non-overlapping frequency bands: In this type of system, various radio node pairs use orthogonal frequency bands for internode synchronization as well as remote sensing operation. Radio nodes operating on non-overlapping frequency bands can track the master reference signals. However, as the phase variations are not accurately tracked (due to significant gaps in the frequency bands), a divergence of the absolute phase is observed. Some notable examples of these systems can be found in [19]–[22].
- Adjacent frequency bands: In this type of system, the participating radio nodes use the same frequency band, but different central frequencies. In comparison to the non-overlapped frequency bands, the absolute phase variation at the frequencies of operation can be tracked accurately.
- Overlapping frequency bands: In this type of system, the radio nodes use the same frequency band and overlapping center frequencies. This type of deployment is capable of accurately tracking the phases at all the frequencies of operation. On the downside, in order to distinguish between the signals from multiple nodes, the signals need to be multiplexed in time or using other multi-user approaches. For example, in-band full-duplex or self-signal cancellation methods show satisfactory results under high signal to noise ratio (SNR) scenarios [23][24].

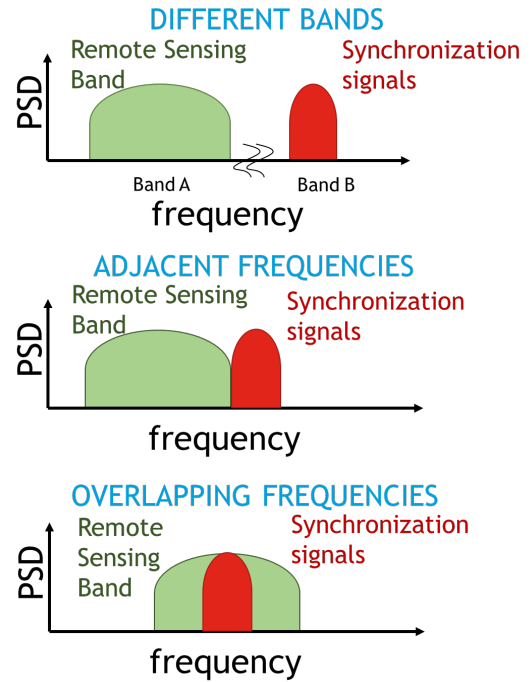

Fig. 2. Classification of system architecture based on the use of the spectrum. Power Spectral Density (PSD) is displayed for the three architectures.

### IV. LITERATURE EXAMPLES OF DISTRIBUTED REMOTE SENSING

The following list contains some relevant examples of distributed remote sensing systems and missions, not necessarily using the same synchronization approach presented in this work:
- Tandem-X: In the Tandem-X mission, which operates in X-band, two Synthetic Aperture Radar (SAR) satellites follow a close formation with the distance between them varying up to 100 meters [21][22]. Tandem-X is capable of performing its operation in both mono-static and bi-static configurations. In bi-static settings, where synchronization between the two satellites is required, one of the satellites operates as the transmitter while the other as the receiver. Further, a round trip synchronization procedure is performed to attain synchronization between the two satellites. A method for synchronization is detailed in [25].

- Low-frequency distributed radio telescope in space (OLFAR) [26][27]: An OLFAR system consists of a space-based low-frequency radio telescope which explores the "dark-ages" of the universe. The OLFAR system will consist of a minimum of fifty satellites and requires synchronization among them. The required synchronization is below one Degree; however, the frequency of target applications is centered at 30MHz.
- Laser Interferometer Space Antenna (LISA) [28]: In the LISA system, laser-based interferometry is used to detect the gravitational waves originating from galactic and extra-galactic sources. By performing a coherent operation, a distance accuracy of approx. 20 ppm can be achieved.
- Gravity Recovery and Climate Experiment (GRACE) [29]: The objective of GRACE was to track changes in the Earth's gravity field. This system consisted of two identical satellites in near-circular orbits separated by approximately 500 km while maintaining a synchronized inter-satellite link. The Gravity Recovery and Interior Laboratory (GRAIL) mission was an analogous mission, but in lunar orbit with two spacecrafts separated by 200 km [30]. In 2018, the GRACE Follow-On (GRACE-FO) mission was launched with very similar characteristics to its predecessor, but with a satellite separation of 220 km [31].

It is worth to mention that phase synchronization is also a harsh requirement for wired round-trip techniques used in radio astronomy applications (e.g. VLA [32]) and wireless techniques developed and applied in space Very Long Baseline Interferometry VLBI (e.g. VSOP [33] or Spektr-R [34]) where the coherent integration time may range from few minutes to hours.

## V. ANALYSIS OF SINGLE FREQUENCY IN-BAND FULL-DUPLEX SYNCHRONIZATION LOOP

The single frequency solution, which consist of two-transponders is represented in Fig. 3. The two satellites (nodes) work in a master and slave (follower) configuration, whereas the master has a high stability reference $u_0(t) = e^{j\theta_0(t)}$, and the follower has a less accurate reference $u_x(t) = e^{j\theta_x(t)}$.

To analyze the effect of non-ideal oscillators, we include the frequency and phase noise as the output of the oscillators $u_m(t) = e^{j\theta_m(t)}$, and the reference $u_0(t)$. The goal of the synchronization algorithm is to have the same phase at the beamforming clock reference in each node. This phase is not a static parameter, and it is affected by the communication channel too. For that reason, it is needed a distributed phase synchronization loop between both nodes to make the two phases $\theta_{\text{bf1}}(t)$ and $\theta_{\text{bf2}}(t)$ equal, despite the phase of the oscillator $\theta_x(t)$ and the phase rotation introduced by the channel (The subscript of $\theta_{\text{bf1}}(t)$ and $\theta_{\text{bf2}}(t)$ is "**bf**" because these are the phases used for the beamforming operations). Notice that all the phases mentioned before are time-dependent. However, from now on, the variable $t$ will be omitted of the equations for the sake of clearness.

The frequency response of the channel is represented in the scheme as its transfer function $H(f)$ and its Laplace equivalent $H(s)$. It is essential to note that, as a first approximation a single frequency and, hence, a symmetric channel is assumed.

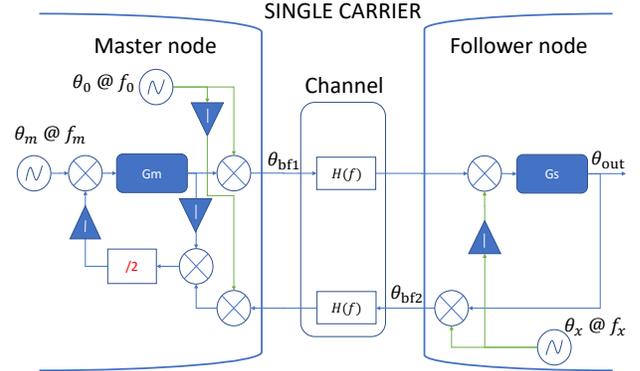

Fig. 3. Schematic diagram for the hypothetical single frequency in-band full-duplex synchronization loop.

The output phase of the master node is determined by

$$\theta_{\text{bf1}} = \theta_0 + G_m(s)\left(\theta_m - \frac{\theta_{\text{bf2}}H(s) - \theta_0 - \theta_g}{2}\right), \quad (1)$$

where $\theta_{\text{bf2}}$ is the reference phase of the follower node, $G_m(s)$ is the s-domain frequency response of the control loop in the master and $\theta_g$ is the phase compensation required to make $\theta_{\text{bf1}} = \theta_{\text{bf2}}$. As can be appreciated, the proposed algorithm is equivalent to a distributed PLL with acquisition and tracking stages determined by the loop equations.

Evaluating $\theta_g = \frac{G_m(s)(2\theta_m - \theta_{\text{bf2}}H(s) - \theta_0)}{2 - G_m(s)}$ and $\theta_{\text{bf2}} = \theta_x(1 + G_s(s)) + \theta_{\text{bf1}}H(s)G_s(s)$ in (2):

$$\theta_{\text{bf1}} = \theta_0 F_{01}(s) + \theta_x F_{x1}(s) - \theta_m F_{m1}(s), \quad (2a)$$

where

$$F_{01}(s) = \frac{G_m(s)}{G_m(s)(1 - 2H^2(s)G_s(s)) - 2}$$

$$F_{x1}(s) = \frac{2G_m(s)H(s)(1 + G_s(s))}{G_m(s)(1 - 2H^2(s)G_s(s)) - 2} \quad (2b)$$

$$F_{m1}(s) = \frac{G_m(s)(2 + G_m(s))}{G_m(s)(1 - 2H^2(s)G_s(s)) - 2}$$

Similarly:

$$\theta_{\text{bf2}} = \theta_0 F_{02}(s) + \theta_x F_{x2}(s) - \theta_m F_{m2}(s) \quad (3a)$$

With:

$$F_{02}(s) = \frac{(3G_m(s) - 2)H(s)G_s(s)}{G_m(s)(1 - 2H^2(s)G_s(s)) - 2}$$

$$F_{x2}(s) = \frac{(G_m(s) - 2)(1 + G_s(s))}{G_m(s)(1 - 2H^2(s)G_s(s)) - 2} \quad (3b)$$

$$F_{m2}(s) = \frac{G_m(s)H(s)G_s(s)(2 + G_m(s))}{G_m(s)(1 - 2H^2(s)G_s(s)) - 2}$$

The the forward and return waveforms use different central frequencies. In this case, the separation can be achieved by applying a frequency offset of $f_{mo}$ to the return waveform. The follower node will generate this offset frequency using its own reference frequency. This frequency offset will inject to the loop a phase ramp of $\theta_x(f_{mo} - f_m)/f_c$. On the other end, the master node will down-convert the received signal with a phase of $\theta_0(f_m - f_{mo})/f_c$. Then, considering that the round trip steady-state phase gain is -1/2, the output phase of the complete distributed phase synchronization loop described by (32a) and (43a) allows $\theta_{bf2}$ to track the changes in $\theta_{bf1}$. However, there is a crucial limitation in this design. It is very challenging to implement the full-duplex link with the continuous transmission and reception of a single carrier frequency as is required for this solution. In an actual full-duplex implementation, there will be residual interference between the transmitted and received signal.

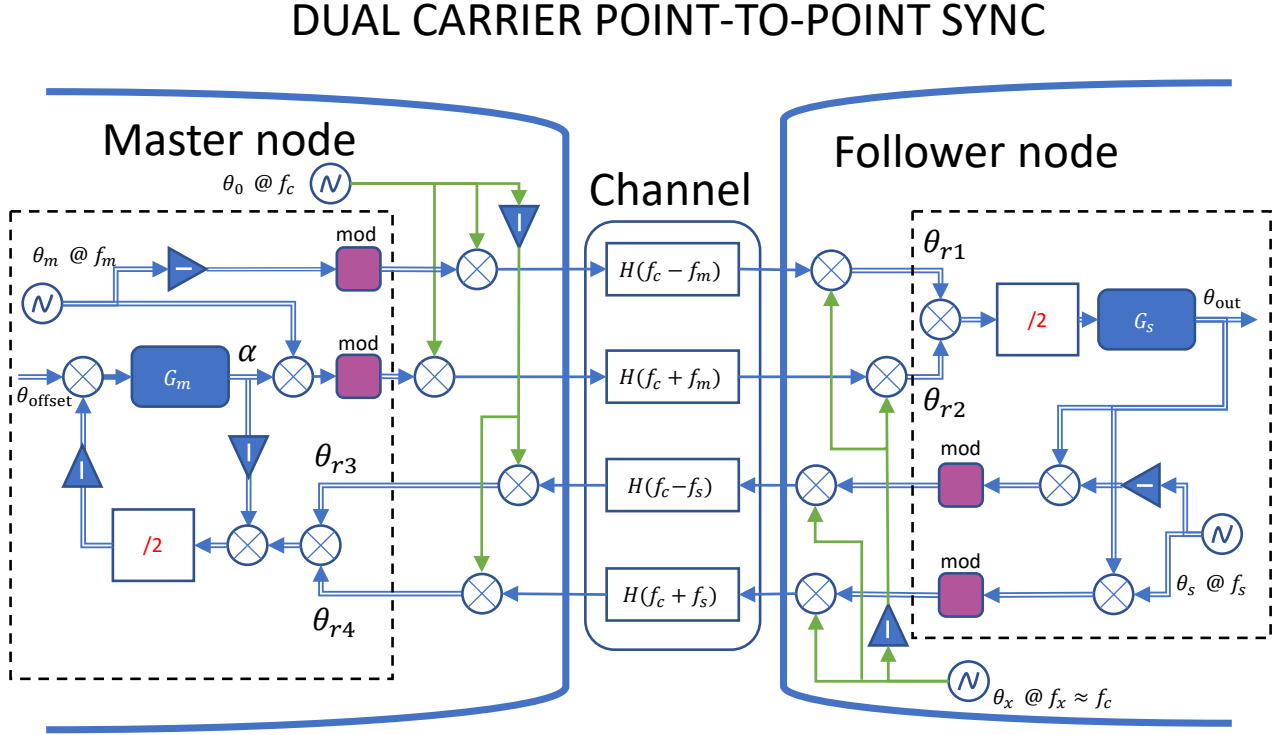

Fig. 4. Schematic diagram for the proposed dual carrier synchronization loop. The dashed line represents the functions implemented digitally. Not shown in the scheme is the beamforming phase obtained as $\theta_{bf} = \theta_{out} + \theta_x$. The purple blocks are the modulators. The demodulators are not explicitly drawn. The double compound lines represent complex variable (capable to express negative frequencies), and the single lines represent real variables.

A solution to overcome these limitations is presented in the next Section.

## VI. PROPOSED DUAL-CARRIER SYNCHRONIZATION LOOP

In the previous Section, the feasibility of a remote phase synchronization loop has been analyzed using the same carrier frequency. It is found that the main limitation for the implementation of the full-duplex loop is the ability to separate incoming and outgoing signals to detect the propagation phases and be able to compensate them.

Let us consider a hypothetical modified single-carrier scheme (per direction), as the one explained before, in which loop will incur an unsolvable error of $-(\theta_x - \theta_0)(f_{mo} - f_m)/2f_c$. This phase offset, which varies over time, might be mitigated by reducing the difference $(f_{mo} - f_m)$ up to the limit where the two carriers start to overlap. Additionally, the asymmetry in the round-trip loop (by using $f_{mo}$ and $f_m$) will substantially augment the phase noise and phase drift injected by RF and microwave components in the loop and the phase offset of the transmission media. Therefore, symmetrical schemes would be preferable in order to address these challenges.

This Section describes a synchronization scheme between a master and a tracker (also known as follower or slave) using two frequency carriers per direction of propagation in a symmetrical fashion. In this scheme, the midpoint of the frequencies of the incoming signals is equal to the midpoint of the outgoing signals. Signal diversity by means of spread-spectrum orthogonal sequences could have been a solution. However, a prior temporal synchronization is necessary in order to cancel their mutual crosstalk, and this is considered out of the scope of this work. Fig. 5 shows a frequency plan of the proposed point-to-point synchronization scheme.

In the proposed dual-carrier scheme, the synchronization signals are placed at the edges of a central frequency $f_c$, which can be the central frequency used for the remote sensing operation. Fig. 5 shows a simplified block diagram of the proposed dual-carrier synchronization scheme. The scheme has the same objective of the single carrier synchronization loop explained in Section V, which is to transfer a very stable frequency reference from the master node toward a follower node.

The master node generates two reference frequency signals from an ultra-stable reference oscillator, which can be an atomic clock. One of these signals is the RF local oscillator used to upconvert the transmitted signals with the frequency $f_c$, and phase $\theta_o$ as shown in Fig. 4. The other signal is the frequency offset $f_m$ which is a complex sinusoidal generated in the digital domain. The master node transmits towards the follower node two modulated carriers, one with a frequency offset $-f_m$ with respect to the central frequency $f_c$, and another one with a frequency offset of $\alpha$ obtained from the whole compensation bidirectional loop.

The follower node receives these two signals after the channel propagation, with phases as $\theta_{r1}$ and $\theta_{r2}$ in Fig. 4. The follower node uses an RF local oscillator with a frequency of $f_x$ trying to make it as close as possible to $f_c$ using a frequency source which can be much less stable than the master oscillator. In order to achieve this goal, the follower node demodulates the transmitted stream, and recover the impingent two phases, and then makes an average and tracks it with the transfer function $G_s$ to obtain the phase

$$\theta_{\text{out}} = \left(\frac{\theta_{r1} + \theta_{r2}}{2}\right) G_s \qquad (5)$$

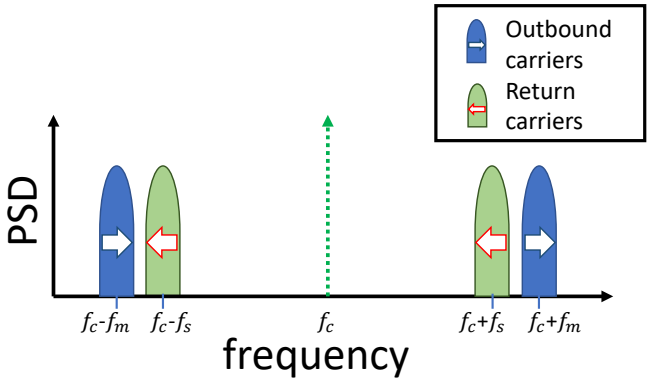

Fig. 4. Frequency plan for the dual-carrier point-to-point synchronization loop. The frequency $\pm f_m$ is the offset with respect to the main carrier $f_c$ of the synchronization signals sent by the master, similarly $\pm f_x$ is the offset of the return signals produced by the follower node.

This output is used to generate two modulated feedback waveforms which are transmitted back to the master node using the frequency offsets $-f_s$ and $f_s$.

A similar process is performed in the master node to generate the loop compensation phase $\alpha$. The impingent modulated carriers, with phases $\theta_{r3}$ and $\theta_{r4}$ are demodulated, (this is not explicitly drawn in Fig. 4 for all the signals). The phases $\theta_{r3}$ and $\theta_{r4}$ of the two received modulated carriers are used to get

$$\alpha = \left(\theta_{\text{offset}} - \frac{\theta_{r3} + \theta_{r4} - \alpha}{2}\right) G_m \qquad (6)$$

Using this compensation phase the control loop cancels the propagation-induced phase fluctuation. As a consequence, the beamforming phase $\theta_{\text{bf}}$, at the follower node will tend to be equal to the phase of the master local oscillator $\theta_0$, with

$$\theta_{\text{bf}} = \theta_{\text{out}} + \theta_x \cong \theta_0 \qquad (7)$$

### A. Transfer function analysis

A transfer function analysis is required to perform closed-loop design in the complete scheme since the output phase is controlled by the master phase and that the channel delays are compensated.

For the analysis, the channel is modeled by a delay $\tau$, idem for the four carriers and, a phase offset $(\theta_1, \theta_2, \theta_3, \theta_4)$ which is the propagation phase wrapped in modulo $2\pi$.

Additionally, the received signal is assumed to be affected by an additive white Gaussian noise. These additive noises are assumed to be statistically independent since the carrier frequencies are different and non-overlapped. Some correlated effects may appear, for instance, due to a poor filtering of the LNA power supply. However, the consideration of these implementation aspects is out of the scope of the present work.

Fig. 6 depicts the channel model used for the phase compensation scheme.

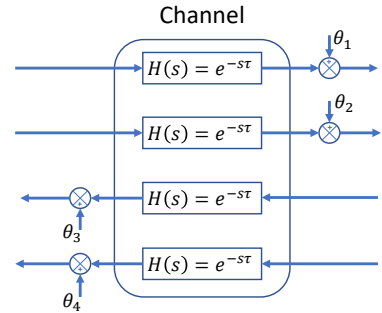

Fig. 5. Equivalent channel model for the dual-carrier phase synchronization scheme. The transfer function is modeled by a delay $e^{-s\tau}$ in the Laplace transform.

The transfer function in the Laplace transform domain of the complete master node block is given by

$$G_c(s) = \frac{\alpha(s)}{\theta_{r3}(s)} = \frac{\alpha(s)}{\theta_{r4}(s)} = \frac{-0.5\, G_m(s)}{1 - 0.5\, G_m(s)}. \qquad (8)$$

This transfer function is used to obtain the phase transfer function from the master oscillator phase $\theta_0$ to the slave recovered phase reference $\theta_{\text{out}}$. This relation is found with the two signals transmitted to the follower node and retrieved in the phases $\theta_{r3}$ and $\theta_{r4}$.

$$\left.\frac{\theta_{out}(s)}{\theta_0(s)}\right|_{\theta_x=0} = \frac{H(s)G_s(s)}{1 - G_c(s)G_s(s)H^2(s)} \quad (9)$$

On the other side, the transfer function from the phase of the follower node oscillator $\theta_x$ to the $\theta_{out}$ is

$$\left.\frac{\theta_{out}(s)}{\theta_x(s)}\right|_{\theta_0=0} = \frac{(H^2(s)G_c(s) - 1)G_s(s)}{1 - G_c(s)G_s(s)H^2(s)} \quad (10)$$

Finally, the transfer function from the phases $\theta_x$ and $\theta_0$ to the beamforming phase $\theta_{bf}$ are found to be

$$\left.\frac{\theta_{bf}(s)}{\theta_0(s)}\right|_{\theta_x=0} = \frac{\theta_{out}(s)}{\theta_0(s)}, \quad (11)$$

and,

$$\left.\frac{\theta_{bf}(s)}{\theta_x(s)}\right|_{\theta_0=0} = \frac{\theta_{out}(s)}{\theta_x(s)} + 1$$
$$= \frac{1 - G_s(s)}{1 - G_c(s)G_s(s)H^2(s)}. \quad (12)$$

We propose to use second-order transfer functions for the PLLs of $G_m(s)$ and $G_s(s)$ because of known benefits, such as tracking capabilities and dynamic response. However, other kinds of loops may be further studied. The second-order transfer functions are described, as usual, by a natural oscillation frequency and a damping factor. Fig. 7-Fig. 9 show the Bode plots for the phase transfer functions in the complete system for the inputs $\theta_0(s)$ and $\theta_x(s)$ for different natural frequencies and damping factors. For these plots $\zeta_m$ is the damping factor of the loop $G_m$ in the master and $\zeta_s$ is the damping factor of the tracking loop $G_s$ in the follower node. The channel delays are assumed to be negligible for the computation of the bode plots. However, these delays may be included in a stability analysis of the system.

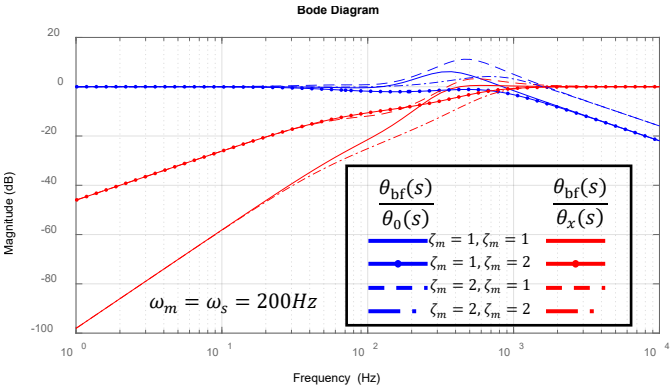

Fig. 6. Frequency response for the synchronization loop for the $\theta_0(s)$ and $\theta_x(s)$ inputs for different damping factors $\zeta_m$ and $\zeta_s$. Here the natural frequency in both the master and the follower is 200 Hz.

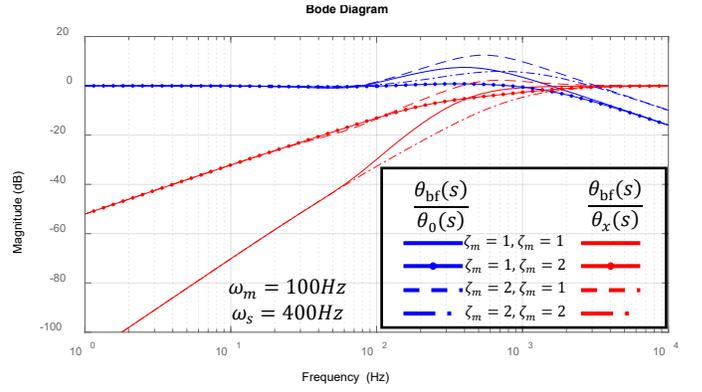

Fig. 7. Frequency response for the synchronization loop for the $\theta_0(s)$ and $\theta_x(s)$ inputs for different damping factors. $\zeta_m$ and $\zeta_s$. Here the natural frequency in the master is $\omega_m = 100$ Hz and the natural frequency in the follower is $\omega_s = 400$ Hz.

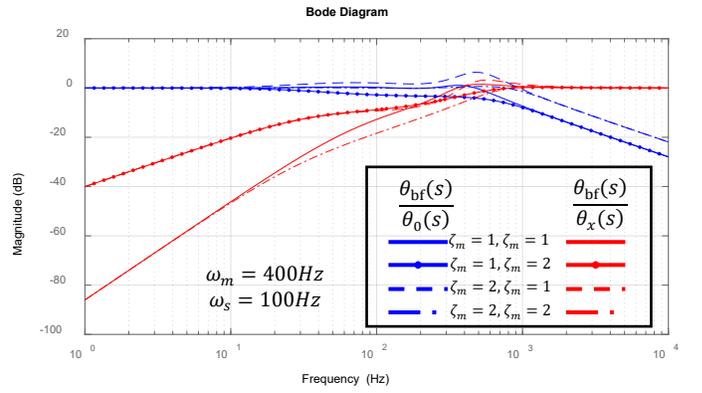

Fig. 8. Frequency response for the synchronization loop for the $\theta_0(s)$ and $\theta_x(s)$ inputs for different damping factors. $\zeta_m$ and $\zeta_s$. Here the natural frequency in the master is $\omega_m = 400$ Hz and the natural frequency in the follower is $\omega_s = 100$ Hz.

### B. Stability and Delay margin analysis

After finding the expression for the transfer function of the synchronization loop, the following step is to make a stability analysis to determine the scenarios and margins in which the system can operate. The system will be stable if all poles of the closed-loop transfer function remain in the left half of the s-plane. This stability is ensured if the magnitude of the open-loop transfer function $G_c(s)G_s(s)H^2(s)$ fulfills the bode stability criterion [35].

The loop stability analysis does not have a gain margin since the open-loop gain is locked because all its factors have unitary magnitude. $G_c(s)$ and $G_s(s)$ are one-to-one tracking PLLs that are designed to be stable). The only external parameter that will have effects in the loop stability is the transport delay $H(s)$. Fig. 10 shows a plot of the delay margin (the maximum acceptable delay before the system loss stability) as a function of the loop natural frequencies, $\omega_m$ and $\omega_s$ having that. $\omega_m = \omega_s$, for $\zeta_m$ and $\zeta_s$ equal to one. This plot shows that the delay margin decreases as the natural frequency increases. Here it can be seen that for the very atypical case of a natural frequency of 1 MHz, the maximum round-trip transport delay is $\tau = 0.23\ \mu s$, which corresponds to a distance of 34 meters in vacuum propagation.

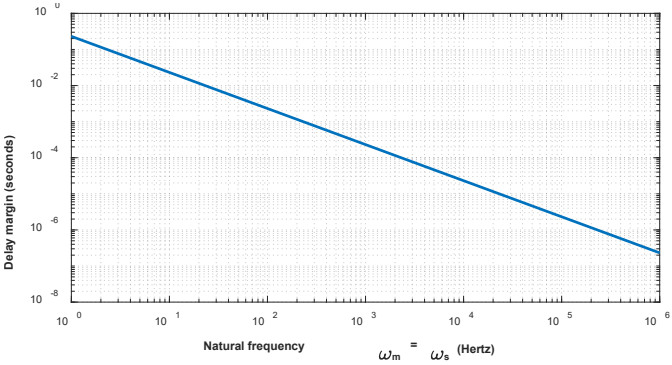

Fig. 9. Delay margin for the synchronization loop as a function of the natural frequencies $\omega_m = \omega_s$. The plot is obtained for $\zeta_m$ and $\zeta_s$ equal to one.

Table I. Simulation parameters.

| Parameter | a |
|---|---|
| Standard | DVB-S2X Super-Frame-Format 2 |
| Baudrate | 8MHz |
| Pilot duration (used) | 32 symbols (out of 36 in standard) |
| Inter-Pilots period | 956 symbols |
| Super-Frame duration | 612540 symbols |
| Pilots modulation | QPSK |
| Pilot sequence type | Walsh-Hadamard (on top of the superframe scrambler) |
| Master offset frequency $f_m$ | 50 MHz |
| Follower offset frequency $f_s$ | 40 MHz |
| Central carrier frequency $f_c$ | 2200 MHz |

## VII. SIMULATIONS OF DUAL-CARRIER SYNCHRONIZATION LOOP

We simulate the proposed dual-carrier synchronization loop using Matlab to validate it in a realistic scenario. The simulations implement the PLLs from the master and follower nodes and interconnect them using the scheme of Fig. 4.

Each PLL consists of a numerically-controlled oscillator, a phase discriminator, and a loop controller in the digital domain. The simulations use the numerology of a DVB-S2X [36] satellite communications using the super-frame format 2. The phase tracking loop in the master and in the follower is performed over the Pilot fields. As seen in Fig. 11, the Inter-Pilots period is 956 symbols.

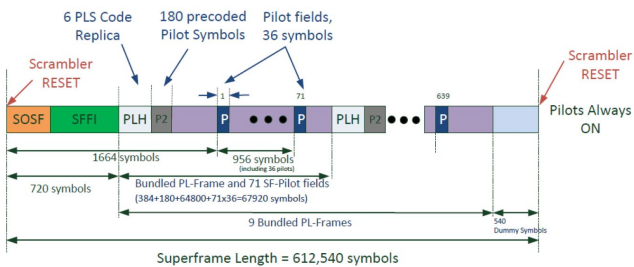

Fig. 10. Super-Frame structure from the DVB-S2X standard [36]. The P fields are used for phase synchronization. The payload symbols can be used for data exchange between the two nodes.

Previous timing and coarse frequency acquisition in the receivers are assumed to be obtained from the frame structure. We made this selection due to the characteristic offered by the standard, such as the periodical pilot sequence used for synchronization. However, the proposed mechanism is general for any kind of coherent communication schemes. The objective of the simulations is to validate the model and verify that a very stable oscillator reference can be transferred to a remote node by canceling out the propagation-induced phase fluctuations. For this purpose, the simulation has the capability to include a phase noise mask for the oscillators in master and follower nodes.

Empirical models based on measurements suggest that the oscillator phase noise PSD can be described as a sum of power-law processes $h_\alpha |f|^\alpha$ with $\alpha \in \{-4, -3, -2, -1, 0\}$ [37] with an additional Gaussian segment near to the carrier [38]. However, for the sake of simplicity, the phase noise is modeled here by a two-state model proposed in [39], which includes a frequency walk, a phase-walk, and a white noise component. The accuracy of this approximation for low-cost oscillators was demonstrated in [40]. Fig. 12 shows a diagram of the model used to reproduce the phase noise behavior in the master and follower nodes.

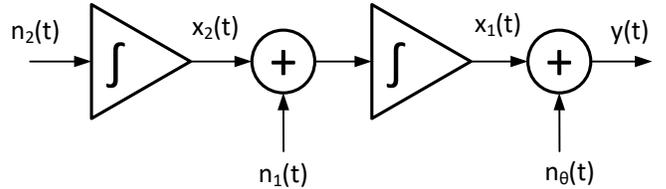

Fig. 11. Two-state oscillator phase noise model. The inputs $n_\theta(t), n_1(t)$ and $n_2(t)$ are real white Gaussian noise processes with their correspondent variances adapted for the required noise mask.

Fig. 13 shows a phase noise plot obtained from a realization of the system in Fig. 12 for around ~501 seconds. The phase noise process is generated using a clock of 8 MHz, and turning the integrators into accumulators in the digital domain.

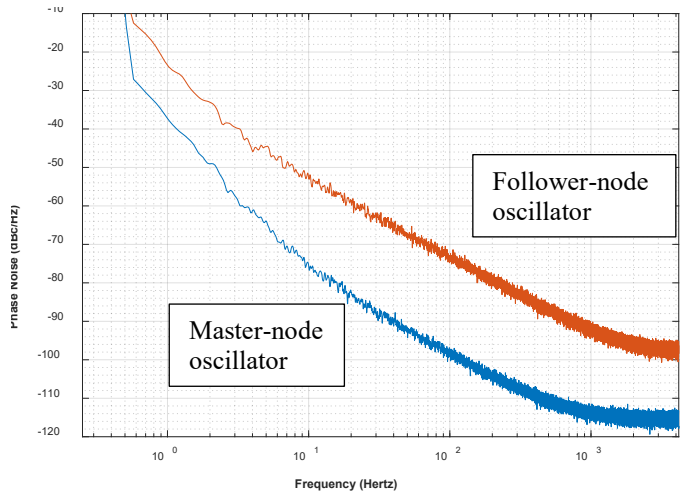

Fig. 12. Estimation of phase noise spectrum in the RF oscillators at 2.2 GHz in the master node and the follower node. These responses are obtained from scaling-up two different qualities 10 MHz clock references. The spectrum is computed using the windowing described in Fig. 14.

The configuration parameters of the two-state oscillator phase noise model were selected according to the phase noise mask described in Table II.

Table II. Phase noise values for the master and follower reference oscillators at 10MHz.

| Phase Noise (dBc/Hz) At 10MHz | Master - Node Oscillator | Follower- Node Oscillator |
|---|---|---|
| 1 Hz | -85 | -70 |
| 10 Hz | -125 | -100 |
| 10 KHz | -160 | -140 |

For the phase noise estimation, the samples are decimated by a factor of 956. A time series of around 501 seconds is used to obtain a total of $2^{22}$ decimated samples. The spectrum is obtained using the periodogram for 32 non-overlapped FFT blocks of $2^{17}$ samples. The master clock phase noise parameters are derived from the specifications of a commercially-available Low Noise Chip Scale Atomic Clock [41]. The follower node phase noise parameters assume any commercial-grade OCXO user in radio applications. The phase noise mask obtained in the RF oscillators shown in Fig. 13 is the reference oscillator multiplied by a 2200/10 scale.

The phase noise spectrum estimation is obtained by windowing the stream of samples by a Dolph-Chebyshev function shown in Fig. 14.

realistic assumption since the channel is assumed to be reciprocal.). Therefore, we assume without any loss of generality that the SNR of the system is the SNR of the received signals in the master node and in the follower node.

The simulations are performed in the digital domain after the pulse compression decimation. Then, the decimated sampling frequency is 8 MHz/956 = 8.36 kHz. The SNR of the received signal has a compression gain, from the 32 samples used in one Pilot field equivalent to 15 dB. The transfer functions in the decimated discrete domain are obtained from the digital implementation of the second-order PLLs, which consist of a loop controller with two integrators. The integrators are replaced by its discrete equivalent using the bilinear transformation $s = 0.5 t_s (z+1)/(z-1)$. Where $t_s$ is the sampling frequency, and $z$ is the z-transform variable. The transfer function for the loop controller in the Laplace domain is

$$Y(s) = (2\zeta\omega + \omega^2/s)/s, \qquad (13)$$

which is a function of generic damping factor $\zeta$ and natural frequency $\omega$. The phase discriminator is implemented with the "angle" function from Matlab and the Numerically Controlled Oscillator (NCO) with the exponential function.

Fig. 15 shows the phase noise for the $\theta_{bf}$ phase output in the follower node for different values of SNR and loop bandwidths $\omega_m$ equal to $\omega_s$. The damping factor for these plots is set to 1.

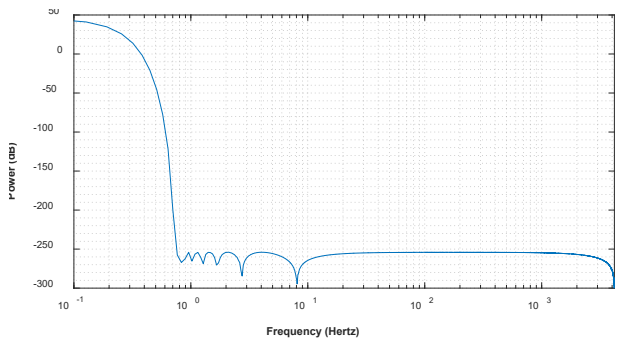

Fig. 13. Power response of the Dolph-Chebyshev window used for the phase noise visualization. The window has a duration of $2^{17}$ samples and it is designed to have a peek-to-sidelobe rejection of 300 dB.

The dynamic system described in Fig. 4 for the proposed dual-frequency synchronization system is simulated under realistic thermal noise conditions in the receivers. The signals in the receivers of the two nodes get to be $r1' = r1 + n_{r1}$, $r2' = r2 + n_{r2}$, $r3' = r3 + n_{r3}$, and $r4' = r4 + n_{r4}$, where $r1, r2, r3,$ and $r4$ are the received signals, and $n_{r1}, n_{r2}, n_{r3}$, and $n_{r4}$, are assumed to be independent additive complex circularly symmetric white Gaussian noise random variables, with the same variance.

Under this simplification, it is assumed that the two nodes transmit the same power in the four carriers and that the four of them are affected by the same channel attenuation (this is a very

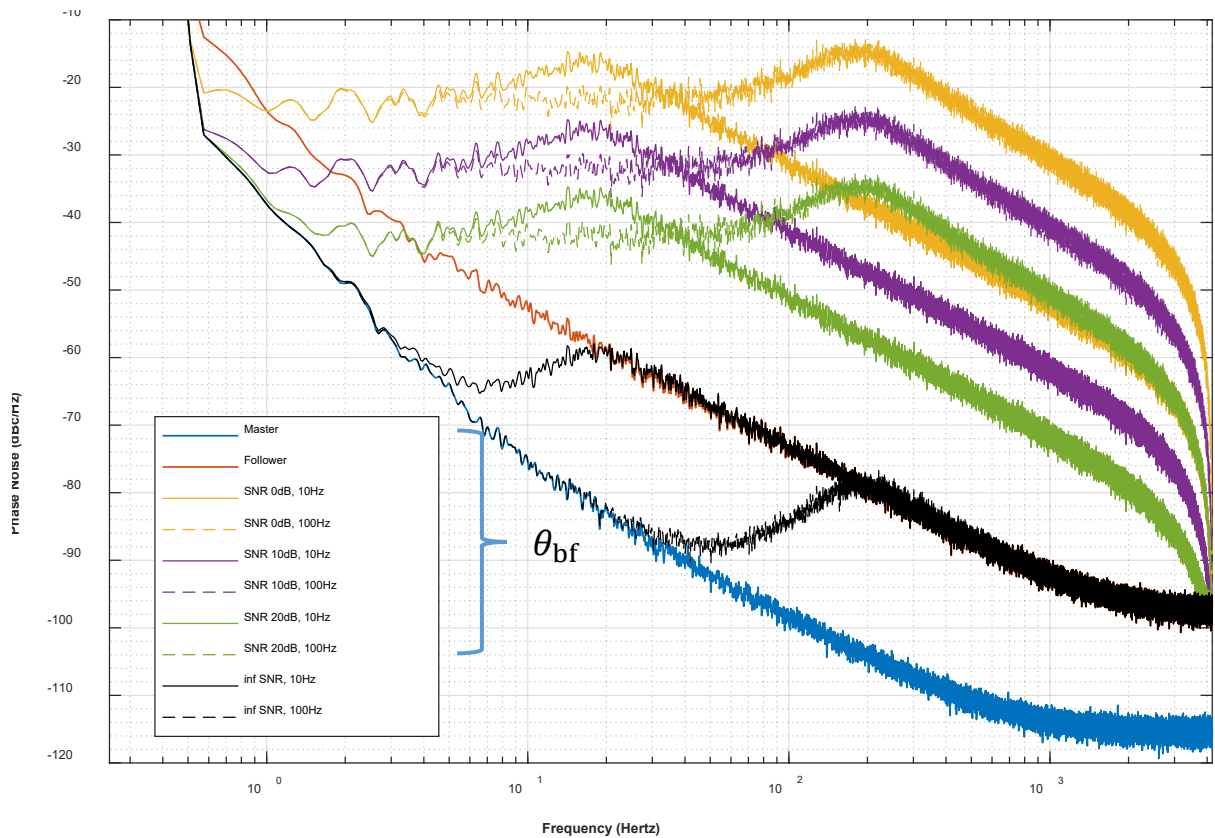

Fig. 14. Estimated Phase noise spectrum for Beamforming phase $\theta_{bf}$, (which is the output of the dual carrier remote phase synchronization system) for different SNRs and loop bandwidth. Both damping factors $\zeta_m$ and $\zeta_s$ are set to one. The phase noise for $\theta_{bf}$ is estimated with the same procedure that for the one of the oscillators in the master and follower node, as is shown in Fig. 4. It is, 32 FFT blocks of $2^{17}$ samples using the window of Fig. 14.

It can be seen how the phase noise of $\theta_{bf}$ tries to reach the phase of the master oscillator for low frequencies. It is observed that the contribution of the Additive White Gaussian Noise (AWGN) inside the noise bandwidth is equivalent to the SNR value, plus the 15 dB of compression gain, plus 6dB coming from the fact that the loop is dual-carrier.

The following figures assess the capabilities of the dual synchronization loop by looking into the difference $\theta_{bf} - \theta_0$ during a given time period, in this case, 120 seconds. Fig. 16 shows the values of this difference for a loop with 0dB SNR. Fig. 17 repeat the plots for a loop with 10dB SNR, and Fig. 18 does it for 20dB SNR.

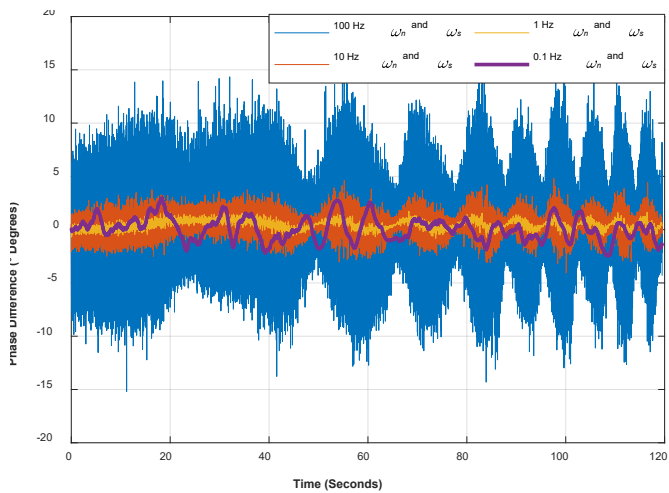

Fig. 15. Time response of the phase difference $(\theta_{bf} - \theta_0)$ for an SNR of 0 dB and different loop bandwidths ($\omega_n = \omega_s$). The damping factors are fixed to one. The sampling rate is 8.36 kHz.

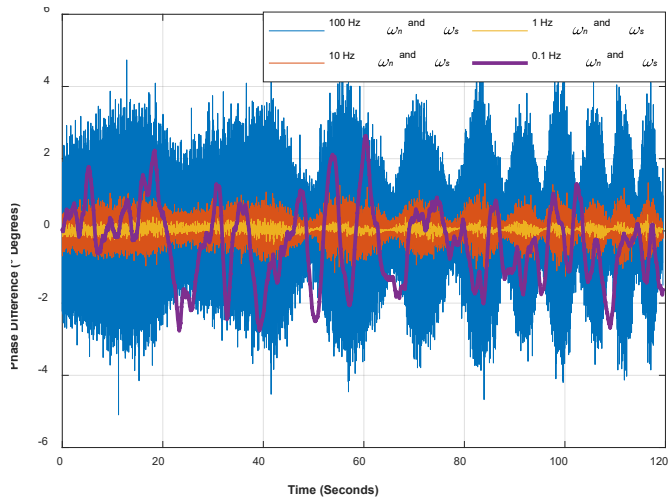

Fig. 16. Time response of the phase difference ($\theta_{bf} - \theta_0$) for an SNR of 10 dB and different loop bandwidths ($\omega_n = \omega_s$). The damping factors are fixed to one. The sampling rate is 8.36 kHz.

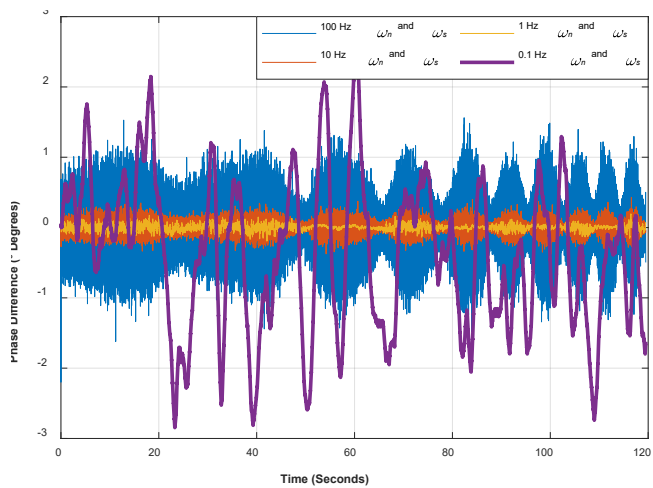

Fig. 17. Time response of the phase difference ($\theta_{bf} - \theta_0$) for an SNR of 20dB and different loop bandwidths ($\omega_n = \omega_s$). The damping factors are fixed to one. The sampling rate is 8.36 kHz.

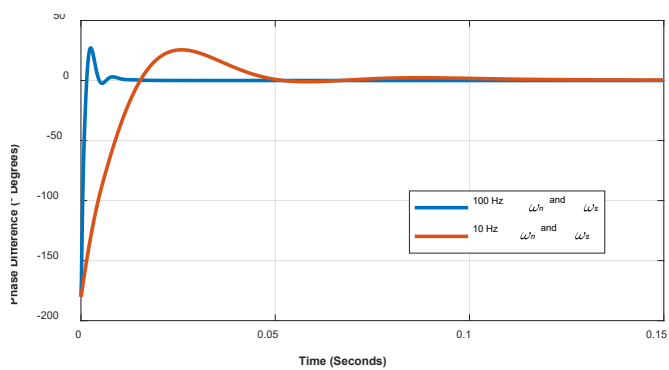

Fig. 18. Time response of the phase difference ($\theta_{bf} - \theta_0$) to an initial phase offset in the follower node oscillator of 180 Degrees. Two loop bandwidths are evaluated: 10 and 100 Hz ($\omega_n = \omega_s$). There is no AWGN in the simulations. The damping factor is 1.

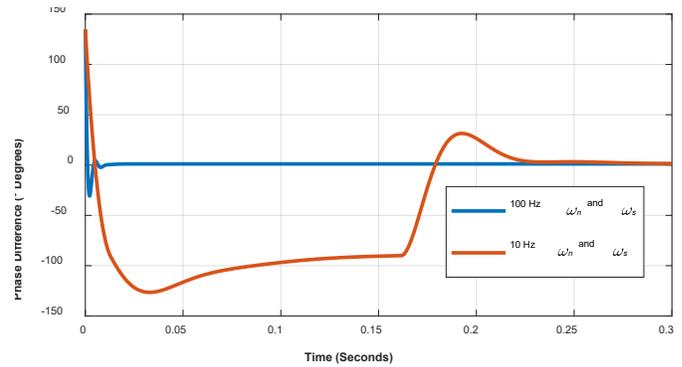

Fig. 19. Time response of the phase difference ($\theta_{bf} - \theta_0$) to a frequency offset in the follower node oscillator of 50 Hz for two different loop bandwidths. There is no AWGN in the simulations. The damping factor is 1.

Fig. 21 and Fig. 22 show a time response of the loop phase difference $\theta_{bf} - \theta_0$ for a constant Doppler shift of 1 Hz, for two different SNR of the loop. One with 10 dB SNR and the other on with infinite SNR. Fig. 21. shows how the phase drift that comes with the Doppler shift does not affect the phase performance of the synchronization loop considerably for the observed time.

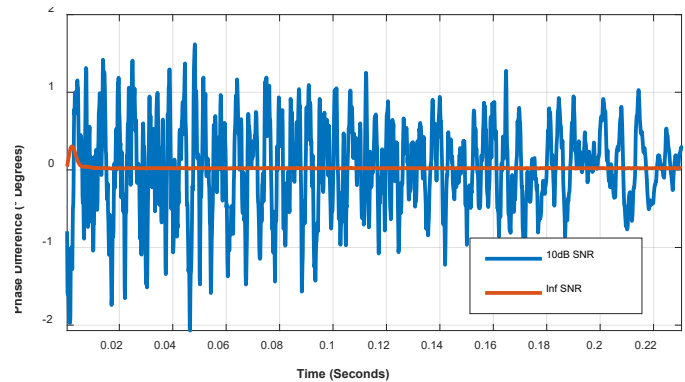

Fig. 20. Time response of the phase difference ($\theta_{bf} - \theta_0$) to a Doppler shift of 1 Hz due to a relative movement between the master and the follower node. Two plots are shown, one with an SNR of 10 dB and another one with infinite SNR. For both plots, the bandwidths are set to 100 Hz ($\omega_n = \omega_s$).

However, it can be observed that a severe Doppler can generate a constant phase offset. It can be roughly approximated to $\Delta\theta = 4\Delta f * \zeta_m\zeta_s/\omega_m\omega_s$ radians, for a second-order tracking PLLs.

There is another effect produced by a relative movement between the two nodes, and it is a 90° phase ambiguity in the loop. Fig. 22 shows how the phase difference $\theta_{bf} - \theta_0$ jumps 90° when the phase of the frequency drift at 1 Hz reaches the 90° limit.

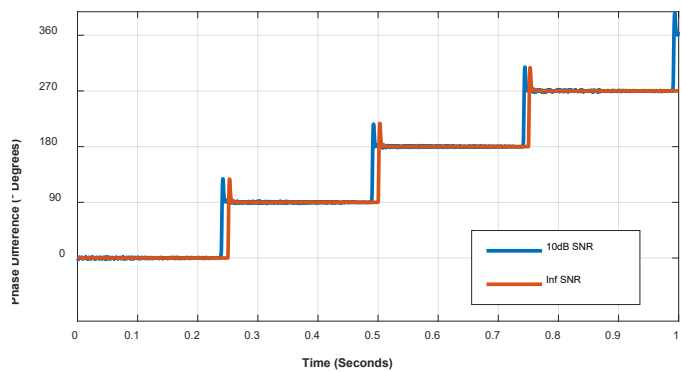

Fig. 21. Time response of the phase difference ($\theta_{bf} - \theta_0$) to an unbounded Doppler shift of 1 Hz due to a relative movement between the master and the follower node. Two plots are shown, one with an SNR of 10 dB and another one with infinite SNR. For both plots, the bandwidths are set to 100 Hz ($\omega_n = \omega_s$). A 90° ambiguity is observed due to the Doppler shift.

The curves of Fig. 22 are obtained from simulations with and without additive noise. It can be seen that the curve with noise jumps around the 90-degree step, with some variability. The ambiguity in the complete loop is a result of the by-two phase dividers in the system, each adding a natural ambiguity of 180°. A mechanism to deal with strong and continuous Doppler shifts will be of interest. It can be resolved to obtain the sign of all of the received and demodulated signals and predicting a wrapping effect. It was also interesting to study this in conjunction with the timing control mechanisms, and an unwrapped channel state estimation.

## VIII. CONCLUSIONS

This work described a method to synchronize the carriers of two remote clock sources to be used in distributed and cooperative remote sensing applications. The paper also studies the possible architectures for the implementation of such systems in spaceborne platforms.

Unlike the prior work in this area, the approach described in this paper allows to have a precise phase synchronization without frequency drifts by using a set of carriers in which the midpoint of the frequencies of the incoming signals is equal to the midpoint of the outgoing signals. This midpoint may be used as the central frequency for the final application and is kept clean by the synchronization loop. Additionally, the carrier of the synchronization loop appears in the same band, in carriers relatively close to each other, separated, for example, by a few MHz. This carrier allocation allows using the same RF front end shared by the synchronization carrier and by the remote sensing application.

We propose the use of conventional and standardized communications links for the implementation of the round-trip remote phase synchronization loop. In this way, the payload symbols can be used for data exchange between the two nodes. The performance of the phase synchronization system was investigated, and results show that the proposed phase synchronization system is effective even in noisy scenarios. For the design of the system, the loop bandwidth selection is a trade-off between the noise and the channel variations response. Results also demonstrate how the performance of the system behaves as a function of the receiver noise, the variations in the phase noise of the follower oscillator, and to Doppler shifts.

It is observed that for the Doppler shift or fast variations in the channel phase, the loop suffers a 90° ambiguity. These kinds of ambiguities are common in remote-phase synchronization loops, which usually contain frequency dividers. For this reason, future work should address the solution of the phase ambiguity in the received streams.

It is worth to mention that the authors have an on-going hardware prototyping of the presented synchronization loop during the year 2020.

## REFERENCES

[1] G. Krieger and A. Moreira, "Spaceborne bi- and multistatic SAR: Potential and challenges," *IEE Proc. Radar, Sonar Navig.*, vol. 153, no. 3, pp. 184–186, Jun. 2006.

[2] P. Lopez-Dekker, H. Rott, P. Prats-Iraola, B. Chapron, K. Scipal, and E. De Witte, "Harmony: an Earth Explorer 10 Mission Candidate to Observe Land, Ice, and Ocean Surface Dynamics," in *IGARSS 2019 - 2019 IEEE International Geoscience and Remote Sensing Symposium*, 2019, pp. 8381–8384.

[3] J. Matar, M. Rodriguez-Cassola, G. Krieger, P. Lopez-Dekker, and A. Moreira, "MEO SAR: System Concepts and Analysis," *IEEE Trans. Geosci. Remote Sens.*, pp. 1–12, 2019.

[4] Z. Yu, J. Chen, C. Li, Z. Li, and Y. Zhang, "Concepts, properties, and imaging technologies for GEO SAR," in *MIPPR 2009: Multispectral Image Acquisition and Processing*, 2009, vol. 7494, p. 749407.

[5] A. M. Guarnieri, A. Broquetas, F. Lopez-Dekker, and F. Rocca, "A geostationary MIMO SAR swarm for quasi-continuous observation," in *2015 IEEE International Geoscience and Remote Sensing Symposium (IGARSS)*, 2015, pp. 2785–2788.

[6] A. Monti Guarnieri, A. Broquetas, A. Recchia, F. Rocca, and J. Ruiz-Rodon, "Advanced Radar Geosynchronous Observation System: ARGOS," *IEEE Geosci. Remote Sens. Lett.*, vol. 12, no. 7, pp. 1406–1410, Jul. 2015.

[7] A. K. Sugihara El Maghraby, A. Grubisic, C. Colombo, and A. Tatnall, "A Novel Interferometric Microwave Radiometer Concept Using Satellite Formation Flight for Geostationary Atmospheric Sounding," *IEEE Trans. Geosci. Remote Sens.*, vol. 56, no. 6, pp. 3487–3498, Jun. 2018.

[8] A. K. Sugihara El Maghraby, A. Grubisic, C. Colombo, and A. Tatnall, "A Novel Interferometric Microwave Radiometer Concept Using Satellite Formation Flight for Geostationary Atmospheric Sounding," *IEEE Trans. Geosci. Remote Sens.*, vol. 56, no. 6, pp. 3487–3498, Jun. 2018.

[9] A. K. S. El Maghraby, A. Grubisic, C. Colombo, and A. Tatnall, "A Hexagonal Pseudo-polar FFT for Formation-Flying Interferometric Radiometry," *IEEE Geosci. Remote Sens. Lett.*, vol. 16, no. 3, pp. 432–436, Mar. 2019.

[10] W. J. Blackwell, "Technology Evolution to Enable High-Performance Cubesat Radiometry Missions," *IGARSS 2019 - 2019 IEEE Int. Geosci. Remote Sens. Symp.*, pp. 5078–5081, Jul. 2019.

[11] A. J. Camps and C. T. Swift, "A two-dimensional Doppler-radiometer for earth observation," *IEEE Trans. Geosci. Remote Sens.*, vol. 39, no. 7, pp. 1566–1572, Jul. 2001.

[12] H. Park and Y.-H. Kim, "Microwave motion induced synthetic aperture radiometer using sparse array," *Radio Sci.*, vol. 44, no. 3, p. n/a-n/a, Jun. 2009.

[13] A. K. S. El Maghraby, H. Park, A. Camps, A. Grubisic, C. Colombo, and A. Tatnall, "Phase and Baseline Calibration for Microwave Interferometric Radiometers Using Beacons," *IEEE Trans. Geosci. Remote Sens.*, pp. 1–12, Feb. 2020.

[14] L. Feng, Q. Li, and Y. Li, "Imaging with 3-D Aperture Synthesis Radiometers," *IEEE Trans. Geosci. Remote Sens.*, vol. 57, no. 4, pp. 2395–2406, 2019.

[15] A. K. S. El Maghraby, H. Park, A. Camps, A. Grubisic, C. Colombo, and A. Tatnall, "An FFT-Based CLEAN Deconvolution Method for Interferometric Microwave Radiometers With Spatially Variable Beam Pattern," *IEEE Geosci. Remote Sens. Lett.*, 2020.

[16] S. Jayaprakasam, S. K. A. Rahim, and C. Y. Leow, "Distributed and Collaborative Beamforming in Wireless Sensor Networks:

Classifications, Trends, and Research Directions," *IEEE Commun. Surv. Tutorials*, vol. 19, no. 4, pp. 2092–2116, 2017.

[17] J. Thompson M. *et al.*, "Phase stabilization of widely separated oscillators," *Antennas Propagation, IEEE Trans.*, vol. 16, no. 6, pp. 683–688, Nov. 1968.

[18] J. C. Merlano-Duncan, J. Querol, A. Camps, S. Chatzinotas, and B. Ottersten, "Architectures and Synchronization Techniques for Coherent Distributed Remote Sensing Systems," *IGARSS 2019 - 2019 IEEE Int. Geosci. Remote Sens. Symp.*, pp. 8875–8878, Jul. 2019.

[19] J. C. Merlano Duncan, "Phase synchronization scheme for very long baseline coherent arrays," Universitat Politècnica de Catalunya, 2012.

[20] J. C. Merlano-Duncan, J. J. Mallorquí, and P. López-Dekker, "Carrier phase synchronisation scheme for very long baseline coherent arrays," *Electron. Lett.*, vol. 48, no. 15, p. 950, 2012.

[21] H. Fiedler *et al.*, "The TanDEM-X mission: an overview," in *2008 International Conference on Radar*, 2008, pp. 60–64.

[22] M. Younis, R. Metzig, and G. Krieger, "Performance Prediction of a Phase Synchronization Link for Bistatic SAR," *IEEE Geosci. Remote Sens. Lett.*, vol. 3, no. 3, pp. 429–433, Jul. 2006.

[23] C. Politis, S. Maleki, J. M. Duncan, J. Krivochiza, S. Chatzinotas, and B. Ottesten, "SDR Implementation of a Testbed for Real-Time Interference Detection With Signal Cancellation," *IEEE Access*, vol. 6, pp. 20807–20821, 2018.

[24] I. R. Pérez *et al.*, "Calibration of correlation radiometers using pseudo-random noise signals," *Sensors*, vol. 9, no. 8, pp. 6131–6149, 2009.

[25] "Method for drift compensation with radar measurements with the aid of reference radar signals," Oct. 2004.

[26] A. Budianu, R. T. Rajan, S. Engelen, A. Meijerink, C. J. M. Verhoeven, and M. J. Bentum, "OLFAR: Adaptive topology for satellite swarms," in *62nd International Astronautical Congress 2011, IAC 2011*, 2011, vol. 9, pp. 7086–7094.

[27] S. P. Chepuri, R. T. Rajan, G. Leus, and A.-J. van der Veen, "Joint Clock Synchronization and Ranging: Asymmetrical Time-Stamping and Passive Listening," *IEEE Signal Process. Lett.*, vol. 20, no. 1, pp. 51–54, Jan. 2013.

[28] "LISA - Laser Interferometer Space Antenna - NASA Home Page." [Online]. Available: https://lisa.nasa.gov/. [Accessed: 07-Jan-2019].

[29] B. D. Tapley, S. Bettadpur, M. Watkins, and C. Reigber, "The gravity recovery and climate experiment: Mission overview and early results," *Geophys. Res. Lett.*, vol. 31, no. 9, p. n/a-n/a, May 2004.

[30] A. J. Evans, "GRAIL Mission," *Encycl. Lunar Sci.*, pp. 1–6, 2017.

[31] F. Flechtner, P. Morton, M. Watkins, and F. Webb, "Status of the GRACE Follow-On Mission," pp. 117–121, 2014.

[32] K. I. Kellermann, E. N. Bouton, and S. S. Brandt, "The Very Large Array," in *Open Skies: The National Radio Astronomy Observatory and Its Impact on US Radio Astronomy*, Cham: Springer International Publishing, 2020, pp. 319–390.

[33] L. I. Gurvits, "Space VLBI: from first ideas to operational missions," *Adv. Sp. Res.*, vol. 65, no. 2, pp. 868–876, Jan. 2020.

[34] Y. Y. Kovalev, N. S. Kardashev, K. I. Kellermann, and P. G. Edwards, "The RadioAstron space VLBI project," in *2014 XXXIth URSI General Assembly and Scientific Symposium (URSI GASS)*, 2014, pp. 1–1.

[35] F. M. Gardner, *Phaselock Techniques*, 2nd ed. John Wiley & Sons, Inc., Hoboken, NJ., 1979.

[36] JTC, "EN 302 307-2 - V1.1.1 - Digital Video Broadcasting (DVB); Second generation framing structure, channel coding and modulation systems for Broadcasting, Interactive Services, News Gathering and other broadband satellite applications; Part 2: DVB-S2 Extensions (DVB-S2X)," 2014.

[37] J. Rutman, "Characterization of Phase and Frequency Instabilities in Precision Frequency Sources: Fifteen Years of Progress.," *Proc. IEEE*, vol. 66, no. 9, pp. 1048–1075, 1978.

[38] A. Chorti and M. Brookes, "A Spectral Model for RF Oscillators With Power-Law Phase Noise," *IEEE Trans. CIRCUITS Syst. Regul. Pap.*, vol. 53, no. 9, 2006.

[39] L. Galleani, "A tutorial on the two-state model of the atomic clock noise," *Metrologia*, vol. 45, no. 6, 2008.

[40] J. Mcneill, S. Razavi, K. Vedula, and D. Richard, "Experimental Characterization and Modeling of Low-Cost Oscillators for Improved Carrier Phase Synchronization."

[41] "Low Noise CSAC (LN-CSAC) | Microsemi." [Online]. Available: https://www.microsemi.com/product-directory/embedded-clocks-frequency-references/4518-low-noise-csac-ln-csac. [Accessed: 30-Mar-2020].

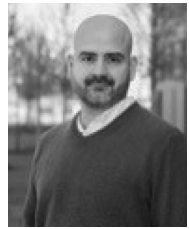

**Juan Carlos Merlano Duncan** (S'09–M'12, SM'20) received the Diploma degree in electrical engineering from the Universidad del Norte, Barranquilla, Colombia, in 2004, the M.Sc. and Ph.D. Diploma (Cum Laude) degrees from the Universitat Politècnica de Catalunya (UPC), Barcelona, Spain, in 2009 and 2012, respectively. His research interests are wireless communications, remote sensing, distributed systems, frequency distribution and carrier synchronization systems, software-defined radios, and embedded systems.

At UPC, he was responsible for the design and implementation of a radar system known as SABRINA, which was the first ground-based bistatic radar receiver using spaceborne platforms, such as ERS-2, ENVISAT, and TerraSAR-X as opportunity transmitters (C and X bands). He was also in charge of the implementation of a ground-based array of transmitters, which was able to monitor land subsidence with sub-wavelength precision. These two implementations involved FPGA design, embedded programming, and analog RF/Microwave design. In 2013, he joined the Institute National de la Recherche Scientifique, Montreal, Canada, as a Research Assistant in the design and implementation of cognitive radio networks employing software development and FPGA programming. He joined the University of Luxembourg since 2016, where he currently works as a Research Scientist in the COMMLAB laboratory working on SDR implementation of satellite and terrestrial communication systems and passive remote sensing systems.

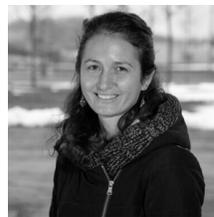

**Liz Martinez Marrero** (S'18) was born in Havana, Cuba, in 1989. She received the M.Sc. degree in telecommunications and telematics from the Technological University of Havana (CUJAE), Cuba, in 2018. She is currently working toward the Ph.D. degree as a Doctoral Researcher at the Interdisciplinary Centre for Security, Reliability, and Trust (SnT) of the University of Luxembourg. Her research interests include digital signal processing for wireless communications, focusing on the physical layer, satellite communications, and carrier synchronization for distributed systems.

During the 37[th] International Communications Satellite Systems Conference (ICSSC2019) she received the Best Student Paper Award.

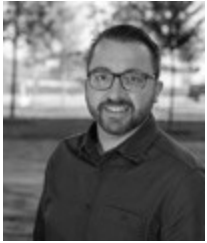

**Jorge Querol** (S'13–M'18) was born in Forcall, Castelló, Spain, in 1987. He received the B.Sc. (+5) degree in telecommunication engineering, the M.Sc. degree in electronics engineering, the M.Sc. degree in photonics, and the Ph.D. degree (Cum Laude) in signal processing and communications from the Universitat Politècnica de Catalunya - BarcelonaTech (UPC), Barcelona, Spain, in 2011, 2012, 2013 and 2018 respectively. His research interests include Software Defined Radios (SDR), real-time signal processing, satellite communications, 5G non-terrestrial networks, satellite navigation, and remote sensing.

His Ph.D. thesis was devoted to the development of novel anti-jamming and counter-interference systems for Global Navigation Satellite Systems (GNSS), GNSS-Reflectometry, and microwave radiometry. One of his outstanding achievements was the development of a real-time standalone pre-correlation mitigation system for GNSS, named FENIX, in a customized Software Defined Radio (SDR) platform. FENIX was patented, licensed and commercialized by MITIC Solutions, a UPC spin-off company.

Since 2018, he is Research Associate at the SIGCOM research group of the Interdisciplinary Centre for Security, Reliability, and Trust (SnT) of the University of Luxembourg, Luxembourg. He is involved in several ESA and Luxembourgish national research projects dealing with signal processing and satellite communications.

He received the best academic record award of the year in Electronics Engineering at UPC in 2012, the first prize of the European Satellite Navigation Competition (ESNC) Barcelona Challenge from the European GNSS Agency (GSA) in 2015, the best innovative project of the Market Assessment Program (MAP) of EADA business school in 2016, the award Isabel P. Trabal from Fundació Caixa d'Enginyers for its quality research during his Ph.D. in 2017, and the best Ph.D. thesis award in remote sensing in Spain from the IEEE Geoscience and Remote Sensing (GRSS) Spanish Chapter in 2019.

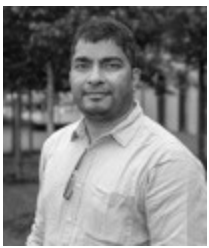

**Sumit Kumar** (S'14-M'19) received his Bachelor of Technology and Master of Science in Electronics & Communication Engineering from Gurukula Kangri University, Haridwar, India (2008) and the International Institute of Information Technology, Hyderabad, India (2014), respectively, and his PhD degree from Eurecom (France) in 2019. His research interests are in wireless communication, interference management and software defined radio.

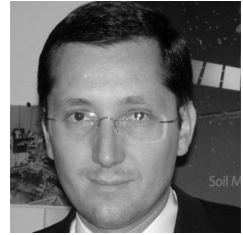

**Adriano Camps** (S'91–A'97–M'00–SM'03–F'11) was born in Barcelona, Spain, in 1969. He received the Graduate degree in telecommunications engineering and the Ph.D. degree in telecommunications engineering from the Universitat Politècnica de Catalunya (UPC), Barcelona, Spain, in 1992 and 1996, respectively. His research interests include microwave remote sensing, with special emphasis in microwave radiometry by aperture synthesis techniques, remote sensing using signals of opportunity (GNSSR), and the development of nanosatellites to test innovative remote sensing techniques.

From 1991 to 1992, he was with the ENS des Télécommunications de Bretagne, Brest, France, with an Erasmus Fellowship. Since 1993, he has been with the Electromagnetics and Photonics Engineering Group, Department of Signal Theory and Communications, UPC, where he was an Assistant Professor first, an Associate Professor in 1997, and a Full Professor since 2007. In 1999, he was on sabbatical leave at the Microwave Remote Sensing Laboratory, University of Massachusetts, Amherst, MA, USA. Since 1993, he has been deeply involved in the European Space Agency SMOS Earth Explorer Mission, from the instrument and algorithmic points of view, performing field experiments and more recently studying the use of Global Navigation Satellite Systems Reflectometry (GNSSR) techniques to perform the sea state correction needed to retrieve salinity from radiometric observations.

Dr. Camps served as a Chair of MicroCal (uCal) 2001, Technical Program Committee Co-chair of IEEE International Geosciences and Remote Sensing Symposium 2007, and Co-chair of GNSS-R 2010. He was an Associate Editor of Radio Science, and he is an Associate Editor of the IEEE Transactions on Geoscience and Remote Sensing and the President-Founder of the IEEE Geoscience and Remote Sensing Chapter in Spain. He received the Second National Award of University Studies, in 1993; the INDRA Award of the Spanish Association of Telecommunication Engineers to the Best Ph.D. in Remote Sensing, in 1997; the Extraordinary Ph.D. Award at the UPC, in 1999; the Research Distinction of the Generalitat de Catalunya, for contributions to microwave passive remote sensing, in 2002; the European Young Investigator Award in 2004; and the Institució Catalana de Recerca i Estudis Avançats Academia Award in 2009 and 2014. Moreover, as a Member of the Microwave Radiometry Group, UPC, he received the first Duran Farell and the Ciutat de Barcelona Awards for Technology Transfer, in 2000 and 2001, respectively, and the "Salvà i Campillo" Award of the Professional Association of Telecommunication Engineers of Catalonia for the most innovative research project for Microwave Imaging Radiometer by Aperture Synthesis/Soil Moisture and Ocean Salinity mission-related activities, in 2004, and the seventh Duran Farell Award for Technological Research, for the work on GNSS-R instrumentation and applications, in 2010.

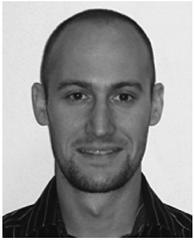

**Symeon Chatzinotas** (S'06–M'09–SM'13) received the M.Eng. degree in telecommunications from the Aristotle University of Thessaloniki, Thessaloniki, Greece, in 2003, and the M.Sc. and Ph.D. degrees in electronic engineering from the University of Surrey, Surrey, U.K., in 2006 and 2009, respectively. He is currently Full-Professor, and the Deputy Head of the SIGCOM Research Group, Interdisciplinary Centre for Security, Reliability, and Trust, University of Luxembourg, Luxembourg, and a Visiting Professor with the University of Parma, Italy. His research interests include multiuser information theory, co-operative/cognitive communications, and wireless network optimization.

He has been involved in numerous research and development projects with the Institute of Informatics Telecommunications, National Center for Scientific Research Demokritos, Institute of Telematics and Informatics, Center of Research and Technology Hellas, and Mobile Communications Research Group, Center of Communication Systems Research, University of Surrey.

Dr. Chatzinotas has more than 300 publications, 5500 citations, and an H-Index of 37, according to Google Scholar. He was a co-recipient of the 2014 Distinguished Contributions to Satellite Communications Award, and Satellite and Space Communications Technical Committee, IEEE Communications Society, and CROWNCOM 2015 Best Paper Award.

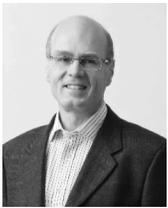

**Björn Ottersten** (S'87–M'89–SM'99–F'04) was born in Stockholm, Sweden, in 1961. He received the M.S. degree in electrical engineering and applied physics from Linköping University, Linköping, Sweden, in 1986, and the Ph.D. degree in electrical engineering from Stanford University, Stanford, CA, USA, in 1989. His research interests include security and trust, reliable wireless communications, and statistical signal processing.

He has held research positions at the Department of Electrical Engineering, Linköping University, the Information Systems Laboratory, Stanford University, the Katholieke Universiteit Leuven, Leuven, Belgium, and the University of Luxembourg, Luxembourg. From 1996 to 1997, he was the Director of research with ArrayComm Inc., a start-up in San Jose, CA, based on his patented technology. In 1991, he was appointed as a Professor of signal processing with the Royal Institute of Technology (KTH), Stockholm, Sweden. From 1992 to 2004, he was the Head of the Department for Signals, Sensors, and Systems, KTH, and from 2004 to 2008, he was the Dean of the School of Electrical Engineering, KTH. He is currently the Director of the Interdisciplinary Centre for Security, Reliability, and Trust, University of Luxembourg.

As a Digital Champion of Luxembourg, Dr. Ottersten acts as an Adviser to the European Commission. He has co-authored journal papers that received the IEEE Signal Processing Society Best Paper Award in 1993, 2001, 2006, and 2013 and three IEEE conference papers receiving Best Paper Awards. In 2011, he received the IEEE Signal Processing Society Technical Achievement Award. He has received the European Research Council advanced research grant twice, from 2009 to 2013 and from 2017 to 2021. He served as an Associate Editor for the IEEE Transactions on Signal Processing and on the Editorial Board of the IEEE Signal Processing Magazine. He is currently an Editor in Chief of the EURASIP Signal Processing Journal and a member of the editorial boards of the EURASIP Journal of Advances Signal Processing and Foundations and Trends of Signal Processing. He is a Fellow of the EURASIP.